\documentclass[reprint,aps,prx]{revtex4-2}
\usepackage{graphicx} 
\usepackage{amsmath} 
\usepackage{amssymb}  
\usepackage{bm}
\usepackage{siunitx}
\usepackage[ISO]{diffcoeff}
\usepackage{float}
\usepackage{url}
\DeclareMathOperator{\diag}{diag}
\DeclareSIUnit\pu{pu}
\begin{document}
\title{Stabilizing Large-Scale Electric Power Grids with Adaptive Inertia}
\author{Julian Fritzsch}
\email{julian.fritzsch@etu.unige.ch}
\author{Philippe Jacquod}
\email{philippe.jacquod@gmail.com}
\affiliation{Department of Quantum Matter Physics, University of Geneva, CH-1211 Geneva, Switzerland}
\affiliation{School of Engineering, University of Applied Sciences of Western Switzerland HES-SO, CH-1950 Sion, Switzerland}

\date{\today}

\begin{abstract}
The stability of AC power grids relies on ancillary services that mitigate frequency fluctuations. 
The electromechanical inertia of large synchronous generators is currently the only resource to absorb frequency disturbances on
sub-second time scales. Replacing standard thermal power plants with inertialess new renewable sources of energy (NRE) therefore jeopardizes grid stability
against e.g. sudden power generation losses. To guarantee system stability and compensate the lack of 
electromechanical inertia in grids with
large penetrations of NREs,  virtual synchronous generators, that emulate conventional generators, have been proposed.
Here, we propose a novel control scheme for virtual synchronous generators, where the provided inertia is large at short times -- thereby absorbing faults 
as efficiently as conventional generators -- but decreases over a tunable time scale to prevent coherent frequency oscillations from setting in.
We evaluate the performance of this adaptive inertia scheme under sudden  power  losses in large-scale transmission grids. We find that it 
systematically outperforms conventional, electromechanical inertia and that it is more stable than previously suggested schemes. 
Numerical simulations  
show how a quasi-optimal geographical distribution of adaptive inertia devices not only  absorbs local faults efficiently, 
but also significantly increases the damping of inter-area oscillations. Our results show that the proposed adaptive inertia control scheme is an excellent 
solution to strengthen grid stability in future low-inertia power grids with large penetrations of NREs. 
\end{abstract}
\maketitle

\section{Introduction}

AC power grids are dynamical systems of oscillators -- representing rotating generators and frequency-dependent loads -- connected to one
another by power lines~\cite{bergenPowerSystemsAnalysis2000,Witthaut22}.
Beyond transporting electricity from producers to consumers, power lines have a dichotomous role. On the one hand, 
they induce  a sufficiently strong coupling to maintain synchrony between oscillators~\cite{Rohden2012,RODRIGUES2016} -- this is necessary for the 
well functioning of the system. 
On the other hand, they propagate frequency waves which may imperil that synchrony~\cite{Kettemann2016,Haehne2019,pagnierInertiaLocationSlow2019,Torres2020,Anvari2020,Zhang2020,Tyloo2023}. 
Neglecting internal degrees of freedom of generators, the frequency dynamics of AC power grids about their synchronous fixed point
is governed by the swing  equations~\cite{machowskiPowerSystemDynamics2020}.
From this set of coupled nonlinear, damped wave equations it is clear that the electromechanical inertia of conventional 
power generators based on rotating machines is a very precious
resource to absorb disturbances on short, sub-second time scales. Simultaneously, inertia favors the long-distance propagation and long-time persistence of 
subharmonic frequency oscillations~\cite{Rogers2012} which, in the worst instances, lead to grid instability and even 
blackouts~\cite{venkatasubramanianAnalysis1996Western2004}. So far, this dichotomous nature of electromechanical inertia has been dealt with 
by introducing frequency control -- in particular the so-called droop control --  which damps frequency oscillations by adapting the power output of generators to the AC grid 
frequency~\cite{machowskiPowerSystemDynamics2020}. Both inertia and droop control are crucial to the stability of AC power grids, 
and as of today, both are provided almost exclusively by conventional generators.

The current push toward decarbonization of the energy sector -- the largest contributor to global carbon dioxide 
emissions~\cite{europeancommission.jointresearchcentre.CO2EmissionsAll2022} -- induces a  strong increase in penetration of new renewable energy sources (NRE) in 
electric power systems, as fossil-fueled power plants are replaced by wind turbines, photovoltaic panels and other NREs. The latter
differ from traditional power plants in at least three fundamental ways. First, they are volatile and have more fluctuating, less predictable power productions.
Second, they are geographically decentralized.
Third, they are connected to the grid via electronic power converters, and thus have an altogether different dynamics from standard power plants with electromechanically 
coupled rotating machines. Increasing the penetration of NRE in power grids therefore implies significant reduction in availability of ancillary services such as 
electromechanical inertia  and droop control, with a simultaneous increase in power fluctuations. Because 
voltage frequency directly reflects imbalances between power generation and
consumption, these power fluctuations result in increased frequency 
fluctuations~\cite{creightonIncreasedWindGeneration,paganiniGlobalPerformanceMetrics2017,Haehne2019,Anvari2020,kerciFrequencyQualityLowInertia2023}. The latter may affect the well-functioning of the grid, as they induce potentially damaging torques on rotating generators, reduce the efficiency of frequency-dependent loads, 
impact industrial processes relying on high current quality or trigger safety devices such as circuit breakers, which 
operate based on fluctuations of voltage frequency. 
Increasing frequency fluctuations while simultaneously reducing resources to mitigate them introduces a number of challenges for grid operators~\cite{milanoFoundationsChallengesLowInertia2018}. 

Various solutions have been proposed to increase the stability of low-inertia grids, one of them being virtual synchronous generators (VSG)~\cite{driesenVirtualSynchronousGenerators2008,zhongSynchronvertersInvertersThat2011}. When 
equipped with electric energy storage, VSGs can emulate the response of inertiaful rotating generators by injecting amounts of additional active power proportional 
to the rate of change of frequency (RoCoF)~\cite{darcoVirtualSynchronousMachines2013,sangVirtualSynchronousGenerator2022}. VSGs being based on
power electronics, their inertia is not physical as in conventional generators, and it can be adapted to the state of the system to improve performances in, e.g., 
mitigating frequency fluctuations.  Such control schemes go by the name of {\it adaptive inertia}.
Ref.~\cite{soniImprovementTransientResponse2013} suggests changing the droop coefficient of the power-frequency control loop with the RoCoF, which increases the virtual inertia provided in large RoCoF events.
Another adaptive inertia strategy tries to minimize the RoCoF and frequency deviation using an on-line optimization of inertia and damping constants~\cite{torresl.SelfTuningVirtualSynchronous2014}.
Other control strategies change both inertia and damping proportional to the RoCoF~\cite{liSelfAdaptiveInertiaDamping2017}.
Refs.~\cite{alipoorPowerSystemStabilization2015,wangAdaptiveControlStrategy2018} propose a bang-bang control strategy focused on 
a fast recovery of the synchronized state -- a method that displays instabilities, however~\cite{kasisStabilityPowerNetworks2021,kasisStabilityPropertiesPower2023}.
Finally, adding a power feedback loop to virtual inertia is proposed in~\cite{liuEquivalentInertiaProvided2022} to keep the RoCoF within predefined bounds.

These investigations considered small grids exclusively.
Accordingly they focused on short time dynamical effects such as the RoCoF and entirely neglect long-range oscillations that adaptive inertia schemes could generate or help sustain.
Having pointed out above the dichotomous nature of inertia, we here 
follow an altogether different approach and investigate a novel adaptive inertia scheme in large-scale power grids. 
Our approach is tailored to explore regimes of clear time-scale separation between fast frequency fluctuations and 
slow coherent oscillations. This approach 
allows us to differentiate the impact of adaptive inertia on 
short-distance RoCoF phenomena vs. long-range inter-area oscillations~\cite{Rogers2012}. 
The novel adaptive inertia control scheme we propose incorporates a driving force, 
increasing inertia at a rate proportional to the absolute value of the RoCoF, and a restoring force 
bringing inertia back to its initial, low value. 
The scheme ensures that inertia (i)  increases fast in cases of a large RoCoF to quickly mitigate its impact, 
and (ii) decreases quickly once short-time fluctuations have been absorbed, 
to re-synchronize the system fast and avoid driving long-range oscillations. 
The performance of this novel adaptive inertia scheme is assessed against several metrics --
frequency and RoCoF based $l_2^2$ norms, resynchronization time, and inertial energy supplied to the grid.
We find that our adaptive scheme systematically  outperforms conventional electromechanical inertia. In particular, it
is able to re-synchronize the grid at an intermediate frequency value, thanks to its absence of dependence on frequency deviation.
Finally, numerical results suggest that adaptive inertia VSGs should be located in peripheral zones in priority.

The paper is organized as follows.
Section~\ref{sec:model} introduces our model.
Section~\ref{sec:stability} investigates the stability of the extended swing equations.
In Section~\ref{sec:applications} we investigate the performance of the extended model on two grids.
First, the IEEE RTS-96 grid is used to demonstrate the impact of the different parameters of the driving and restoring force and to visualize the effect of the adaptive inertia.
Second, we use a model of the European high-voltage power grid to investigate the effect of the adaptive inertia on a strongly-connected large-scale grid.
We emphasize that the impact of the adaptive inertia on the performance is largest when it is located in peripheral areas.
Conclusions and final discussions are given in Section~\ref{sec:conclusions}.


\section{The Model and the Adaptive Inertia Control Scheme\label{sec:model}}

The state of an AC power grid is determined by voltage angles and amplitudes at all of its nodes. At steady-state, these cooordinates
are determined by power flow equations, which combine Ohm's and Kirchhoff's laws~\cite{machowskiPowerSystemDynamics2020,bergenPowerSystemsAnalysis2000}.
A perturbation such as a sudden line opening by a circuit breaker, or a 
fast disconnection of a large generator or load induces a transient voltage dynamics.
Under the standard assumption that there
is a time-scale separation in the dynamics of the voltage angle vs. its amplitude, and focusing on time scales from sub-seconds
to few tens of seconds, voltage amplitudes can be considered constant and one focuses on the voltage angle dynamics. 
This dynamics is different at generator and load nodes. 
Voltage angles at generator nodes are governed by the so-called swing equations. 
In the lossless line approximation -- justified for high voltage grids -- they read~\cite{machowskiPowerSystemDynamics2020,bergenPowerSystemsAnalysis2000}
\begin{equation}\label{eq:swinggen}
    m_i \dot{\omega}_i + d_i \omega_i = P_i - \sum_{j}b_{ij}\sin\left( \theta_i - \theta_j \right),
\end{equation}
where $m_i$ is the inertia of the $i$th generator, $d_i$ its damping parameter, $\theta_i$ its voltage angle, 
$\omega_i=\dot{\theta_i}$ the deviation of its angular frequency from the rated frequency
and $P_i>0$ the generated power. Finally, nodes are coupled by lines with susceptances $\tilde{b}_{ij}$
 and $b_{ij}=\tilde{b}_{ij} V_i V_j$ is the product of that susceptance with the voltage magnitudes (assumed constant) 
at nodes $i$ and $j$.
Renewable energy sources are inertialess and modeled as 
\begin{equation}\label{eq:swingload}
    d_i \omega_i = P_i - \sum_j b_{ij}\sin\left(\theta_i - \theta_j\right).
\end{equation}
We use the structure preserving model~\cite{bergenStructurePreservingModel1981}, according to which the voltage angle dynamics at loads is also
described by Eq.~\eqref{eq:swingload}, where $d_i$ now gives the frequency dependence of the load, and $P_i<0$ is the consumed power.

If all nodes had inertia, the dynamics of voltage angles would be determined by 
a discrete, nonlinear, damped wave equation, Eq.~\eqref{eq:swinggen}. In real power grids, typically one node out of ten 
is a generator node, therefore the wave dynamics is superimposed on a background diffusion equation, Eq.~\eqref{eq:swingload}.
The frequency-dependence constant $d_i$ is however very small for loads, resulting in a large diffusion constant and accordingly a quasi-ballistic 
propagation of voltage angle waves between generator nodes. 

\begin{figure}
    \centering
    \includegraphics[width=\columnwidth]{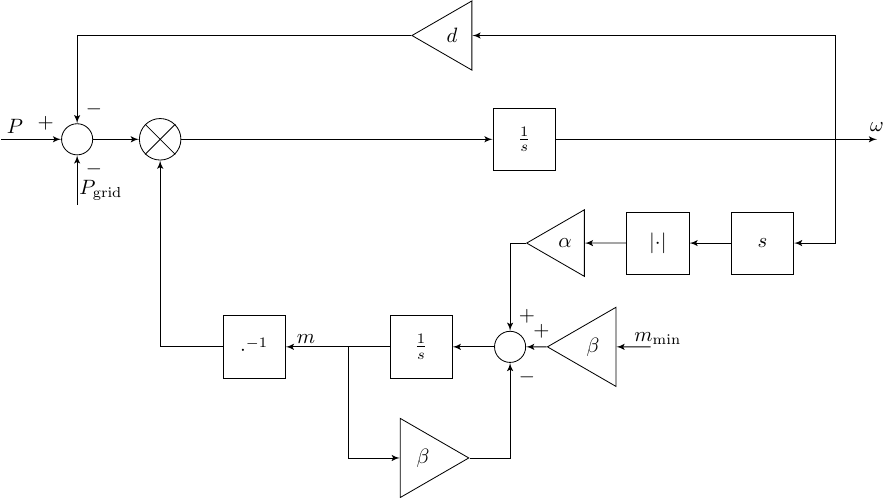}
    \caption{Block diagram of our proposed adaptive inertia scheme. $P_\mathrm{grid}$ is the power flow across the lines connected to the generator. It is given by the sine terms of the power flow equations~\eqref{eq:swinggen}.\label{fig:block}}
\end{figure}

Eqs.~\eqref{eq:swinggen} and \eqref{eq:swingload} give the dynamics in a  power grid with conventional and NRE generators. 
When additionally one has a VSG with adaptive inertia, say at node $i$, Eq.~\eqref{eq:swinggen} is augmented by 
\begin{equation}\label{eq:adaptinertia}
    \dot{m}_i = \alpha_i |\dot{\omega}_i| - \beta_i(m_i - m_{\mathrm{min}, i}) \, ,
\end{equation}
which defines our novel adaptive inertia scheme. It assumes that there is a minimal amount $m_{\mathrm{min}, i} > 0$ of inertia present at all times.
Two control parameters further determine the dynamical evolution $m_i(t)$ as a frequency disturbance $\dot{\omega}_i \ne 0$ hits node $i$. 
First, the gain $\alpha_i > 0$ controls the short-time increase and the maximum amount of inertia to absorb that disturbance. Second, 
$\beta_i > 0$ models a restoring force and controls the rate at which inertia returns to its minimal amount, in order to minimize its contribution
to the further propagation of frequency disturbances.
A block diagram of the proposed control scheme, Eq.~\eqref{eq:adaptinertia}, is shown in Fig~\ref{fig:block}.
An important feature of our adaptive inertia scheme is that, because it depends only on the frequency derivative -- the RoCoF -- 
and not on the frequency itself,  it is able to synchronize at an intermediate frequency value unlike, e.g., the bang-bang scheme which explicitly depends on the frequency.
The two control parameters $\alpha_i$ and $\beta_i$ need to be tuned to optimally absorb large RoCoF values and to make the system return quickly to the steady state.
In Section S1 of the supplemental material~\cite{supp} we give an illustrative example that demonstrates the action of our adaptive inertia method on a single machine infinite bus model.

\section{Stability of the Control Scheme\label{sec:stability}}

Previously proposed adaptive inertia schemes generate instabilities in certain circumstances~\cite{kasisStabilityPowerNetworks2021,kasisStabilityPropertiesPower2023}
and we first show that our scheme does not introduce any such instability.
Eq.~\eqref{eq:adaptinertia} cannot be linearized about the synchronous state as the derivative of $|\dot{\omega}|$ is not defined at $\dot{\omega}=0$.
We therefore introduce a small deadband $\varepsilon \ge 0$ into Eq.~\eqref{eq:adaptinertia} without changing the fixed point,
\begin{equation}\label{eq:adaptive_eps}
    \dot{m}_i = \frac{1}{2}\alpha_i \left( |\dot{\omega}_i + \varepsilon | +|\dot{\omega}_i - \varepsilon | \right) - \beta_i(m_i - m_{\mathrm{min},i}) - \alpha_i\varepsilon,
\end{equation}
where $0 < \varepsilon \ll 1$.
For $\varepsilon\to0$ we recover the original Eq.~\eqref{eq:adaptinertia}.
This mathematical trick allows us to linearize Eqs.~\eqref{eq:swinggen}--\eqref{eq:adaptinertia} about a fixed point with $\omega_i = \delta\omega$, $\theta_i = \theta^0_i + \delta \theta_i$, $m_i = m_{\mathrm{min}, i} + \delta m_i$, $P = P^0_i + \delta P_i$.
Without loss of generality, we assume that there are only generator nodes with adaptive inertia.
The linearized equations then read
\begin{equation}\label{eq:stability}
    \begin{bmatrix}
        \bm{\delta\dot{\theta}} \\
        \bm{\delta\dot{\omega}} \\
        \bm{\delta \dot{m}}
    \end{bmatrix} =
    \begin{bmatrix}
        \bm{0}                          & \bm{1}                          & \bm{0}      \\
        -\bm{M_\mathrm{m}}^{-1}\bm{L} & -\bm{M_\mathrm{m}}^{-1}\bm{D} & \bm{0}      \\
        \bm{0}                          & \bm{0}                          & -\bm{\beta}
    \end{bmatrix}
    \begin{bmatrix}
        \bm{\delta\theta} \\
        \bm{\delta\omega} \\
        \bm{\delta m}
    \end{bmatrix} +
    \begin{bmatrix}\bm{0} \\ \bm{\delta P}\\ \bm{0}\end{bmatrix},
\end{equation}
where $\bm{\delta\theta} = (\delta\theta_1, ... )$, $\bm{\delta\omega} = (\delta\omega_1, ...)$, $\bm{\delta m} = (\delta m_1, ...)$, $\bm{\delta P} = (\delta P_1, ...)$
$\bm{M_\mathrm{m}} = \diag\left(m_{\mathrm{min},i}\right)$, $\bm{D} = \diag\left(d_i\right)$, $\bm{\beta} = \diag\left( \beta_i \right)$ and $\bm{L}$ is the network Laplacian defined as
\begin{equation}
    \bm{L}_{ij} = \begin{cases}
        -b_{ij}\cos(\theta_i^0-\theta_j^0)      & \text{for }i\neq j, \\
        \sum_k b_{ik}\cos(\theta_i^0-\theta_k^0) & \text{for } i = j.
    \end{cases}
\end{equation}
The stability matrix in Eq.~\eqref{eq:stability} is negative semidefinite because (i) its sub-block 
\begin{equation}
    \bm{A} = \begin{bmatrix}\bm{0}                          & \bm{1}                          \\
               -\bm{M_\mathrm{m}}^{-1}\bm{L} & -\bm{M_\mathrm{m}}^{-1}\bm{D}
    \end{bmatrix} 
\end{equation}
is the stability matrix of a system with conventional generators with inertia $m_{\mathrm{min}, i}$ and (ii) the restoring parameters $\beta_i > 0$.
This guarantees the linear stability of our adaptive inertia control scheme.

\section{Numerical Simulations\label{sec:applications}}
We evaluate the efficiency of our adaptive inertia control scheme by investigating the frequency response following a step change in the active power injected by a  
generator. We consider both step changes in the production of conventional generators and of VSGs. We measure the grid response against
four different performance measures, chosen to be sensitive to various dynamical aspects. We consider   
$l_2^2$ norms of the frequency and of the RoCoF
\begin{align}
    l_2^2(\omega)       & = \sum_i\int_0^\infty(\omega_i(t)-\omega_\mathrm{sync})^2\dl t, \label{eq:freq}\\
    l_2^2(\dot{\omega}) & = \sum_i\int_0^\infty\dot{\omega}_i(t)^2\dl t,\label{eq:rocof}
\end{align}
the energy injected into the grid by the inertial response
\begin{equation}
    E_\mathrm{rot} = -\sum_i \int_0^\infty m_i(t)\dot{\omega}_i(t)\dl t \, ,\label{eq:energy}
\end{equation}
as well as the resynchronization time $t_{\mathrm{re}}$. We define the latter as the time it takes to damp frequency deviations down to less than \SI{1}{\milli\hertz} on all nodes.
These four performance measures are sensitive to various dynamical features -- from short-time, fast frequency disturbances, to long-time, large-scale
coherent inter-area oscillations. 
Optimizing our adaptive inertia scheme with respect to all of them -- which is feasible as we find below -- guarantees an overall quasi-optimal disturbance mitigation protocol.

Before presenting numerical results, we qualitatively discuss the influence of the control parameters 
$\alpha_i$ and $\beta_i$ on the performance measures.
Eq.~\eqref{eq:adaptinertia} makes it clear that the large RoCoF following a fault increases the inertia 
until the RoCoF is sufficiently small or the inertia sufficiently large that the second term on its right-hand side dominates, and relaxes the inertia back to its 
minimal value.
The short time behavior is therefore dominantly impacted by $\alpha_i$, and the long term behavior by $\beta_i$.
Next, the frequency performance measure, Eq.~\eqref{eq:freq}, is strongly influenced by long-lived, sustained frequency oscillations, while
the RoCoF performance measure, Eq.~\eqref{eq:rocof}, is mostly determined by the short-time dynamics directly following the fault.
Finally, the resynchronization time is by definition linked to the long time system dynamics.
Putting all this together, we expect the frequency and the resynchronization time performance measures to improve with increasing $\beta_i$, 
whereas the RoCoF performance measure should improve with $\alpha_i$. We furthermore expect that
the injected energy performance measure, Eq.~\eqref{eq:energy}, depends non-trivially on both $\alpha_i$ and $\beta_i$, 
as it explicitly depends both on the RoCoF and the inertia.
The response of the system is stronger when the size $\Delta P$ of the fault increases.
We further expect a scaling behavior where the response remains the same when~\cite{caveat}
\begin{equation}\label{eq:scaling}
    \alpha_i \, \Delta P_i = \text{const}\, .
\end{equation}
This follows from the fact that the RoCoF is given by~\cite{machowskiPowerSystemDynamics2020}
\begin{equation}\label{eq:initrocof}
    \dot{\omega}_i(t=0) = \frac{\Delta P_i}{m_i}
\end{equation}
together with Eq.~\eqref{eq:adaptinertia}.
This scaling is confirmed in Fig.~\ref{fig:faultsize} below.

We next compare the performance measures for networks with and without VSGs.
The networks considered are the IEEE RTS-96 network~\cite{griggIEEEReliabilityTest1999} and \emph{PanTaGruEl}, a model of the synchronous grid of continental 
Europe~\cite{pagnierInertiaLocationSlow2019,tylooKeyPlayerProblem2019}.

\begin{figure}
    \centering
    \includegraphics{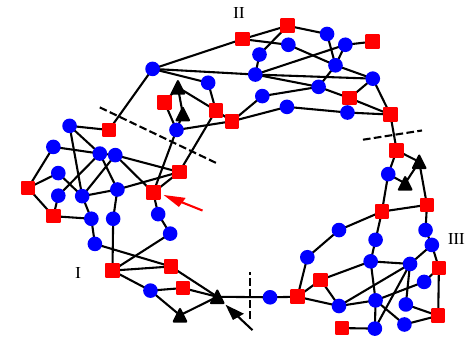}
    \caption{The IEEE RTS-96 network. The red squares are conventional generators, the black triangles are VSGs equipped with the adaptive inertia method, 
    Eq.~\eqref{eq:adaptinertia},
    and the blue circles are load nodes. The black arrow indicates the VSG where the power is changed and the red arrow indicates the conventional generator at which a fault is applied.
    The roman numerals label the different areas and the dashed lines indicate their boundaries.\label{fig:rts96}}
\end{figure}

\subsection{IEEE RTS-96\label{sec:rts}}
We begin by investigating the IEEE RTS-96 test case shown in Fig.~\ref{fig:rts96}.
It is partitioned into three areas with ten generation units each.
The remaining nodes are loads.
In the default case all of the generation units are conventional generators.
For evaluation of our method we promote two of the generation units in each area to VSGs equipped with the adaptive inertia method, Eq.~\eqref{eq:adaptinertia}.
Following the standard of power systems, all physical quantities are expressed in the {\it per unit system} which is obtained by division with some common base unit~\cite{machowskiPowerSystemDynamics2020}.
Static physical quantities such as line capacities and power injections and consumptions are defined in Ref.~\cite{ieeerts96}. 
The inertia of conventional generators is randomly drawn from a uniform distribution, $m_i \in [\SI{0.1}{\pu}, \SI{1,1}{\pu}]$ and each damping corresponds to  a fixed ratio $d_i / m_i \approx 0.3$. 
The chosen inertia range is typical for generators with a maximum production of the order of \SI{100}{\mega\watt} to \SI{1}{\giga\watt}.
For the adaptive generators the initial inertia $m_\mathrm{min}$ is set to one third of the randomly drawn inertia in the default case.
Other parameters are the same for all VSGs.
The fault considered is an instantaneous 25 \% reduction of the power generated by 
the VSG with adaptive inertia marked by a black arrow in Fig.~\ref{fig:rts96}. 
\begin{figure}
    \centering
    \includegraphics{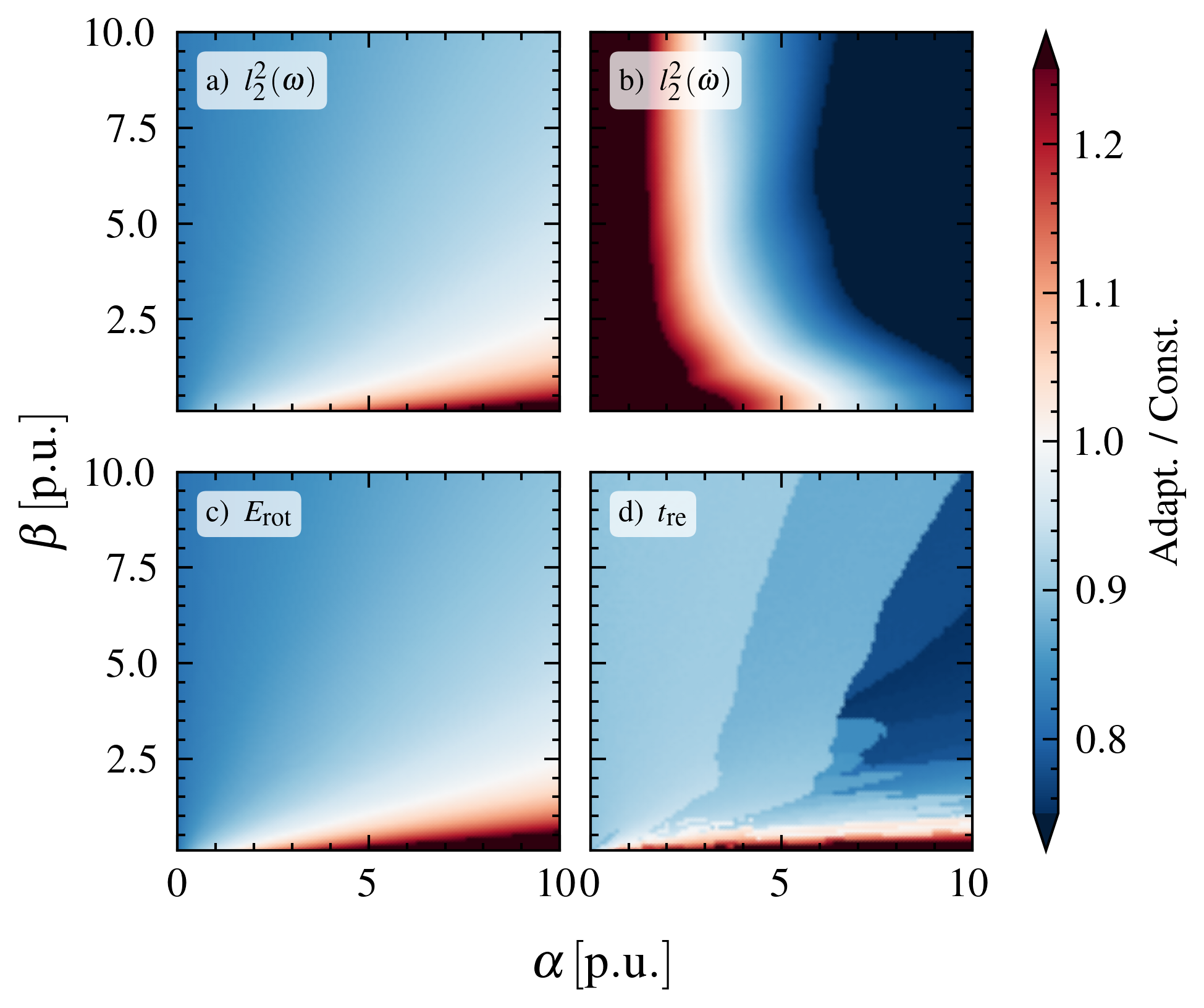}
    \caption{ Dependence of a) the frequency performance measure, Eq.~\eqref{eq:freq}, b) the RoCoF performance measure, Eq.~\eqref{eq:rocof}, c) the injected inertial energy, Eq.~\eqref{eq:energy}, and d) the resynchronization time, $t_{\rm re}$, on the VSG control parameters $\alpha$ and $\beta$ defined in Eq.~\eqref{eq:adaptinertia}. The fault considered is
a change in the power at the generator indicated by the black arrow in Fig.~\ref{fig:rts96}. Color coded are the ratios of the performance measures with 
VSGs over those with only conventional generation. Blue colored areas correspond to the adaptive method performing better than conventional generators.\label{fig:sweep}}
\end{figure}
Fig.~\ref{fig:sweep} shows the ratio of each of the four performance measures with and without VSGs.
Panels a), b) and d) confirm our above expectations that the frequency performance measure 
and the resynchronization time improve with $\beta$, while the RoCoF performance measure
improves with $\alpha$. Panel b) further shows that $l_2^2(\dot{\omega})$ has a weak dependence on $\beta$, which is only visible at small $\beta$,
when generator oscillations following a fault persist longer.
Next, the data in panel a) depend on the ratio $\alpha/\beta$ and not individually on $\alpha$ nor $\beta$. We find that this is the case, because 
fixing $\alpha/\beta$ gives similar time profiles for $m(t)$, with in particular a similar maximal inertia value. Increasing $\alpha$ and $\beta$ while keeping their ratio fixed gives a faster rise
of virtual inertia, which does not affect the long-time dynamics. It therefore leaves $l_2^2(\omega)$ mostly unchanged. It is remarkable that the injected energy performance measure, panel c), 
exhibits the same behavior as $l_2^2(\omega)$. This suggests that it is dominated by long-term dynamical effects.  
Finally, the resynchronization time exhibits a more intricate behavior, with an optimal performance
at large $\alpha$ and medium $\beta$ instead of $\beta \gg \alpha$ as expected. We find that this behavior, however, depends on the fault location and even more on
the network considered.
Fig.~\ref{fig:faultsize} shows the validity of the scaling, Eq.~\eqref{eq:scaling}.
The three subplots look the same when the horizontal axis is scaled with the size of the fault.
Additionally, we find, but do not show, that the performance improves with the number of VSGs in the grid.
Overall, the best global performances are obtained for both $\alpha, \beta \gg 1$. 
Finally, we point out that the performance improves with the number of VSGs in the grid.
\begin{figure}[!ht]
    \centering
    \includegraphics[width=\columnwidth]{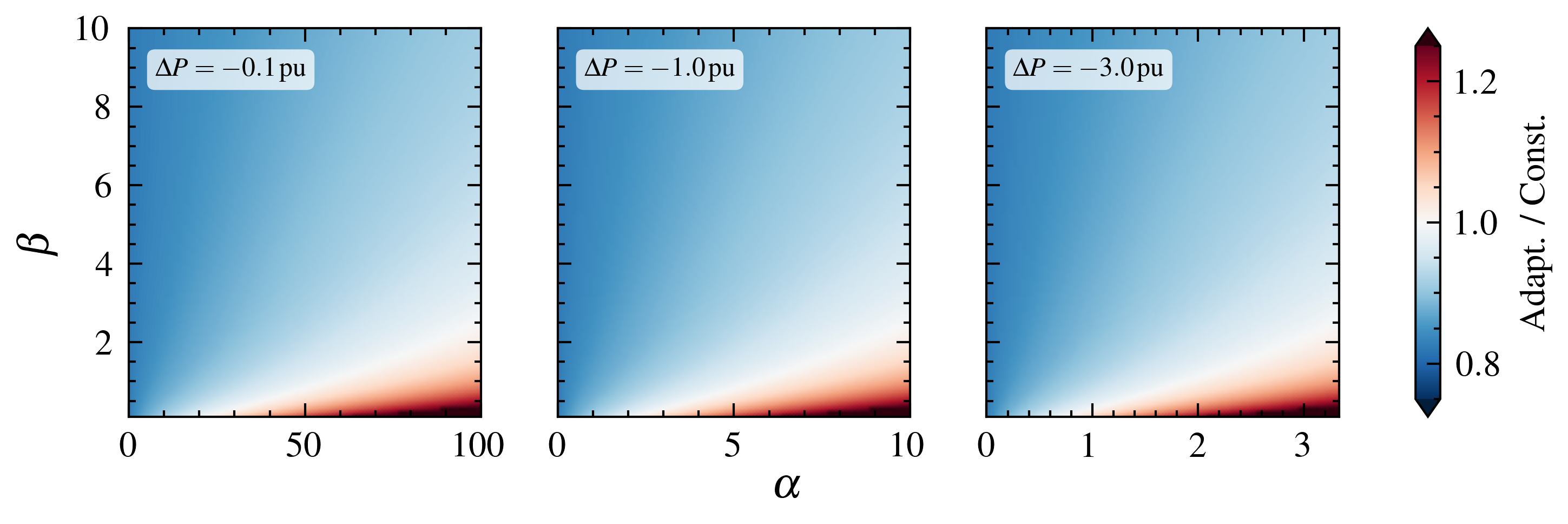}
    \caption{Frequency performance measure, Eq.~\eqref{eq:freq}, for a fault of size $\Delta P = \SI{-0.1}{\pu}$ (left), $\Delta P = \SI{-1.0}{\pu}$ (middle), and $\Delta P = \SI{-3.0}{\pu}$ (right) on the VSG marked by a black arrow in Fig.~\ref{fig:rts96}.
    Color coded are the ratios of the performance measures with VSGs over those with only conventional generation. Blue colored areas correspond to the adaptive method performing better than conventional generators.\label{fig:faultsize}}
\end{figure}

\begin{figure}[!ht]
    \centering
    \includegraphics[width=\columnwidth]{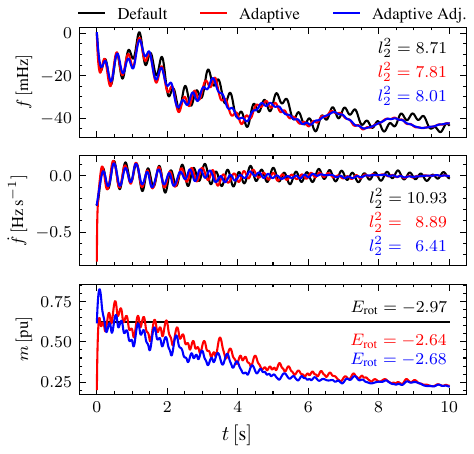}
    \caption{Time evolution of the frequency, $f = \omega/2 \pi$ (top panel), the RoCoF, $\dot{f} = \dot{\omega}/2\pi$ (middle panel), and the inertia (bottom panel) following a change of $\delta P =\SI{-1}{\pu}$ at the VSG indicated by the black arrow in Fig.~\ref{fig:rts96}.
        The black line corresponds to the default case with only conventional generation.
        The red line corresponds to the scenario with six generators being equipped with our adaptive method.
        The blue line corresponds to the adaptive scenario adjusted with an initial inertia of the VSGs set to the value of the default case.\label{fig:rts96comp}}
\end{figure}

In the following we investigate dynamical effects in some more detail for $\alpha = \beta = \SI{5}{\pu}$.
Fig.~\ref{fig:rts96comp} illustrates the dynamics of the faulted generator, for the constant inertia case (black curve) and the case with homogeneous VSGs with $\alpha = \beta = \SI{5}{\pu}$ (red curve). To better connect with AC power engineering, we plot the voltage frequency, $f=\omega/2 \pi$ and frequency RoCoF, $\dot{f}=\dot{\omega}/2\pi$.
First, the amplitude of oscillations of the frequency, as well as their short period components are significantly reduced by the adaptive inertia method. This
is directly reflected in the frequency performance measure and in a reduced resynchronization time, $t_{\mathrm{re},\mathrm{adapt}} = \SI{16.3}{\second} < t_{\mathrm{re},\mathrm{const}} = \SI{18.5}{\second}$. Second, the adaptive inertia also improves the 
RoCoF performance measure significantly, as it quickly leads to smaller, slower oscillations of $\dot{f}$.
Furthermore, once $\dot{f}$ has sufficiently decreased, the lower inertia leads to even less oscillations and a fast recovery of the system in the adaptive scheme.
There is a large, initial RoCoF when the faulted generator is a VSG. This is expected, as  the RoCoF is inversely proportional to the inertia right after the fault, Eq.~\eqref{eq:initrocof}, and the initial inertia $m_{min}$  
in the adaptive case has been chosen at one third of its value in the constant inertia case. 

That feature affects only the dynamics of the faulted generator and is not a problem as long as the latter functions purely on power electronics. 
In cases when $m_{min}$  is provided by physical, electromechanically coupled inertia, or if the VSG is connected to the same bus as some conventional generation,
this large initial RoCoF needs to be mitigated, however. This can be achieved by resetting 
 the inertia to a larger value $m(t \le 0) > m_{min}$ once the frequency stays within some predefined range over a period of time (\emph{e.g.} it does not deviate by more than \SI{0.05}{\hertz} from the synchronized state for \SI{1}{\minute}). This adjusted adapted inertia scheme 
adds the benefit of greatly decreasing the initial RoCoF while not losing the advantage of the decaying inertia. 
In our case setting $m_{i, \mathrm{adapt.}}(t=0) = m_{i,\mathrm{const.}}$ leads to a small penalty on the frequency performance measure and the energy injected while more than 
halving the maximum RoCoF and decreasing the RoCoF performance measure significantly (blue curves in Fig.~\ref{fig:rts96comp}).
Fig.~\ref{fig:rts96comp} finally shows the time evolution of the inertia of the faulted generator. With the adaptive scheme, 
the inertia shoots up right after the fault and then decays back to its minimal value.

\begin{figure}[!ht]
    \centering
    \includegraphics[width=\columnwidth]{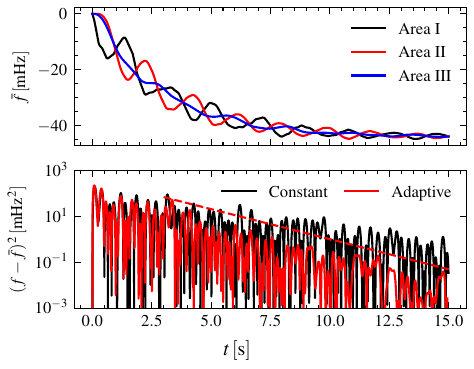}
    \caption{Top panel: Time evolution of the average frequency $\bar{f}$ in each of the three areas of the RTS96 model of Fig.~\ref{fig:rts96}. 
    Inter-area oscillations between area I and II are clearly seen. Bottom panel: Time evolution of the frequency variance around $\bar{f}$ in each area.
    The coherency of each area is enhanced in the presence of VSGs with adaptive inertia (red curve), with an exponential reduction of $(f-\bar{f})^2$ (dashed red line).
\label{fig:inter_intra}}
\end{figure}

We conclude that, for sufficiently large control parameters $\alpha$ and $\beta$, 
our adaptive inertia control scheme outperforms the constant inertia method for all four performance measures. 
Figs.~\ref{fig:sweep} and \ref{fig:rts96comp} make it clear that this occurs, because VSGs perform equally well as conventional inertial generators
at absorbing the short-time RoCoF, but damp frequency oscillations faster. A closer look at how this is done in practice reveals an interesting 
dual action, where VSGs simultaneously fight inter- and intra-area oscillations. The top panel of Fig.~\ref{fig:inter_intra} shows that 
areas I and II  (see Fig.~\ref{fig:rts96}) oscillate against each other -- the fault triggered an inter-area oscillation mode. We find that these modes are more efficiently damped
by VSGs than by conventional generators. Furthermore, VSGs equipped with our adaptive inertia control scheme 
lead to a more homogeneous response and an enhanced intra-area coherency. The effect is quantified by 
the coherency measure
\begin{equation}\label{eq:l2coh}
    l_2^2(\mathrm{coh})=\sum_i\int_0^\infty\left(\omega_i(t) - \bar{\omega}_\mathrm{area}(t)\right)^2\dl t \, , 
\end{equation}
where $\bar{\omega}_\mathrm{area}(t)$ is the mean frequency in the area corresponding to node $i$. The bottom panel of Fig.~\ref{fig:inter_intra} 
shows that the integrand of Eq.~\eqref{eq:l2coh} decays exponentially faster with than without VSGs. Thanks to VSGs, all generators within one area
quickly respond coherently, with their frequency oscillating closer and closer to the area average shown in the top panel. 
We find that the coherence measure for the adaptive case is only \SI{75}{\percent} of the measure for the constant case.
A very interesting finding is therefore that 
the adaptive inertia scheme of Eq.~\eqref{eq:adaptinertia} damps not only inter-area oscillations, but also intra-area, machine-machine oscillations.

Finally, we find, that the situation is similar for a fault on one of the conventional generators.
For a  fault on the generator marked by the red arrow in Fig.~\ref{fig:rts96}, we find that the results for the frequency performance measure as well as the energy and resynchronization time resemble closely panels a), c), and d) of Fig.~\ref{fig:sweep}.
The adaptive case also performs better than the conventional inertia case.
For the RoCoF performance measure, however, the improvement is significantly smaller than the one shown in Fig.~\ref{fig:sweep}.
This is not surprising when the fault is some distance away from the nearest VSG, since the RoCoF is mostly determined by local, short-time dynamics.
These results are shown in Section S2 of the supplemental material~\cite{supp}.

\begin{figure}[ht]
    \centering
    \includegraphics{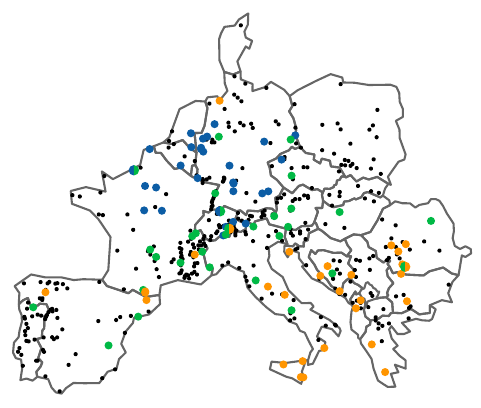}
    \caption{PanTaGruEl model of the synchronous grid of continental Europe. Dots indicate the geographical location of power generators.  
    Black dots correspond to generators that are always conventional machines with electromechanical inertia. 
    Colored dots correspond to VSGs with adaptive inertia in configurations corresponding to central distribution  (blue), peripheral distribution (orange) 
    and homogeneous distribution (green) of VSGs, and to conventional generators otherwise. 
    All colored dots are VSGs in the configuration corresponding 
    to the total distribution.\label{fig:resistance_map}}
\end{figure}

\subsection{PanTaGruEl model of the synchronous grid of continental Europe}

Large-scale power grid models have a clearer time-scale separation and accordingly enable better evaluations of the performance of our 
adaptive inertia control scheme. Therefore we now turn our attention to a continental-size transmission grid.
PanTaGruEl is a model of the synchronous transmission grid of continental Europe.
It consists of 3809 nodes and 7343 lines, encompassing all lines at voltages of 200kV and above~\cite{pagnierInertiaLocationSlow2019,tylooKeyPlayerProblem2019}.
We use a generation dispatch where 448 nodes are generators and promote some of those generators to VSGs with adaptive inertia.
We chose how VSGs are located following three different distributions where 30 VSGs are located centrally, peripherally or homogeneously, 
according to the resistance distance centrality~\cite{Klein1993,tylooKeyPlayerProblem2019}. Details of how this is done are discussed below. 
We further consider a distribution including all 86 different VSGs from these three distributions (four VSGs are included in both the homogeneous and either the central or the peripheral distribution).
Fig.~\ref{fig:resistance_map} shows the location of the VSGs and of the conventional generators for each of these distributions. 

\begin{figure}
    \centering
    \includegraphics[width=\columnwidth]{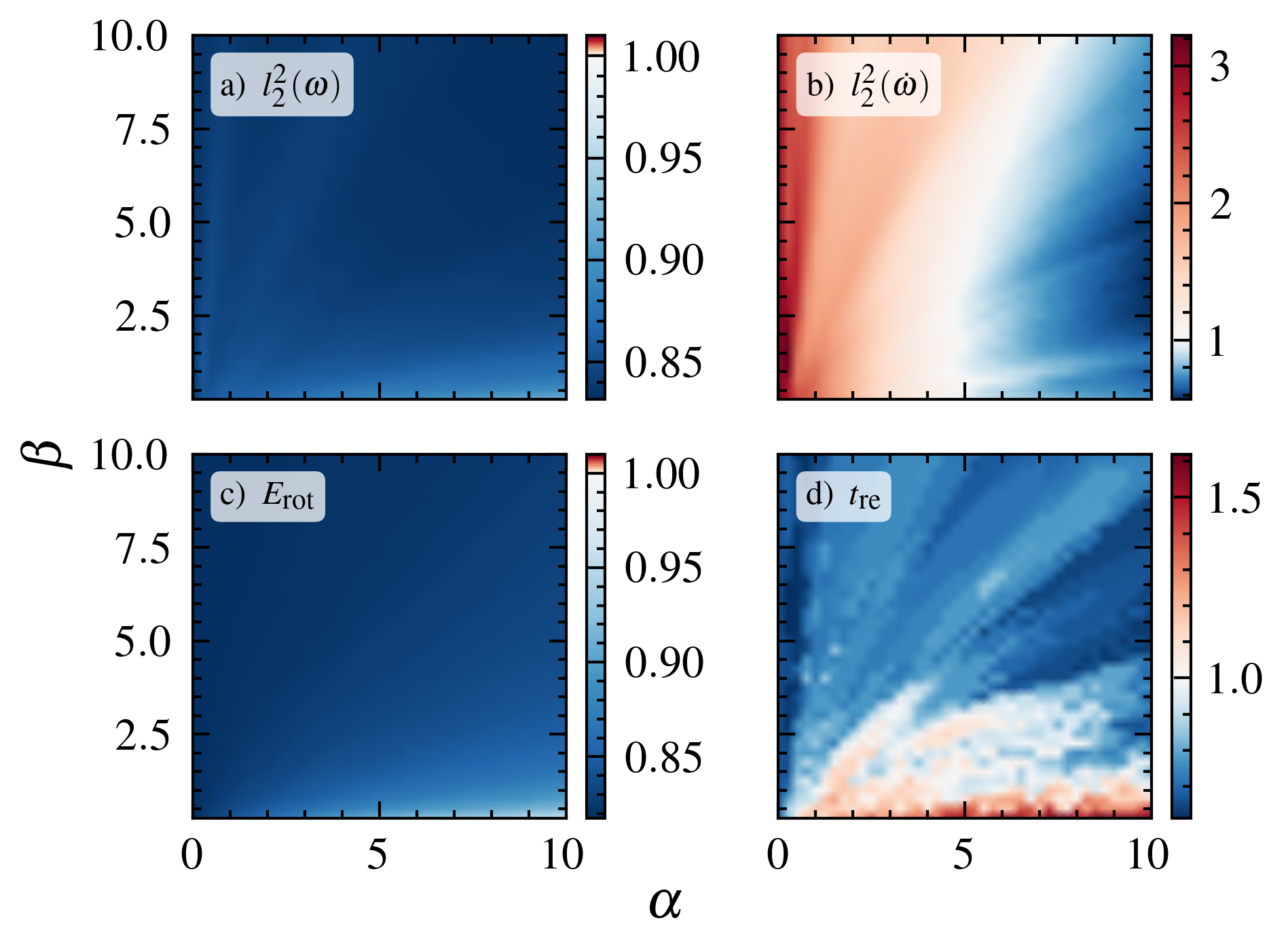}
    \caption{Dependence of a) the frequency performance measure, Eq.~\eqref{eq:freq}, b) the RoCoF performance measure, Eq.~\eqref{eq:rocof}, c) the injected inertial energy, Eq.~\eqref{eq:energy}, and d) the resynchronization time, $t_\mathrm{re}$, on the VSG control parameters $\alpha$ and $\beta$ defined in Eq.~\eqref{eq:adaptinertia}. The fault considered is a \SI{100}{\mega\watt} power loss at a VSG in Switzerland. Color coded are the ratios of the performance measures with VSGs over those with only conventional generation. Blue colored areas correspond to the adaptive method performing better than conventional generators. \label{fig:sweeppanta}}
\end{figure}

The minimum inertia $m_{\mathrm{min}, i}$ of the VSGs is set at one third of their original, constant inertia.
We choose $\alpha = \beta = \SI{10}{\pu}$ which according to Fig.~\ref{fig:sweep} is expected to give excellent performances. 
This is confirmed in Fig.~\ref{fig:sweeppanta} for the PanTaGruEl model, where we show a parameter sweep similar to the one in Sec.~\ref{sec:rts} for a \SI{100}{\mega\watt} fault on a VSG in Switzerland.
In the previous section we focused on the effect of our adaptive method when the power change occurs at one VSG.
In this section we want to investigate the effect on the overall stability of the grid when a fault occurs at a conventional generator.
This is motivated by the much larger scale of PanTaGruEl compared to the RTS-96 grid, and the fact that for practical, financial reasons, only a minority of generators can be promoted to VSGs
in a real, large-scale power grid. 
Of particular interest is to determine if the significantly improved network performances discussed in Section~\ref{sec:rts} persist
when the distance between faulted generation and VSGs grows.
\begin{figure}
    \centering
    \includegraphics{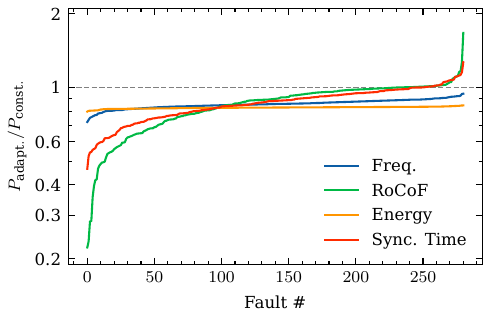}
    \caption{Ratio of the performance measures with constant inertia only vs. with adaptive inertia on 86 VSGs located on the colored
    dots in Fig.~\ref{fig:resistance_map}, on a log-scale.
    Values below the dashed line correspond to our adaptive scheme performing better and values above correspond to the default case with only constant inertia performing better.
    To improve readability, the faults are ordered independently for each curve to make the latter monotonically increasing.\label{fig:pantagruel}}
\end{figure}

\subsubsection{Total distribution of adaptive inertia}

We simulate instantaneous power losses of \SI{100}{\mega\watt} on each of the 281 conventional generators with $P_i^0 \ge \SI{100}{\mega\watt}$ (there are 81 generators that produce less than \SI{100}{\mega\watt} in the used dispatch) and first compare
the conventional inertia case to a grid with a distribution of VSGs on all 86 colored nodes in Fig.~\ref{fig:resistance_map}.
Fig.~\ref{fig:pantagruel} shows the ratio of the four performance measures with and without adaptive inertia VSGs,
 for the 281 individual faults.
Our  adaptive inertia scheme clearly outperforms the constant inertia case almost always:
The frequency performance improves in all cases, while
approximately \SI{17}{\percent} less energy is needed to stabilize the grid. Resynchronization occurs faster for \SI{88}{\percent} of the faults considered.
The situation is less clear when considering the RoCoF performance measure of Eq.~\eqref{eq:rocof} which varies over three orders of magnitude.
This is so because the RoCoF is a local measure: faults on weakly connected generators with low inertia give rise to much larger RoCoF values compared to faults on large power plants in very densely connected areas.
In  \SI{82}{\percent} of all cases, the RoCoF performance measure improves with adaptive inertia.
Generally, it does not improve when the faulted generator lies in a central, well connected area with a high density of inertial generators, 
where RoCoFs are small to start with. For instance,
the two worst cases where the performance measure decreases by more than 20 \% 
correspond to faults whose RoCoF responses are among the 7 smallest ones, out of the 281 considered faults.
We conclude that our adaptive inertia control scheme significantly improves grid performance 
for all faults that may be problematic in the constant inertia case. 

\subsubsection{Geographical distribution of adaptive inertia}

Previous work showed that geographically homogeneous distributions of conventional inertia optimize the grid response against
frequency disturbances~\cite{pagnierOptimalPlacementInertia2019,poollaPlacingRotationalInertia2019}, and a practical question that 
naturally arises is where financially limited resources of adaptive inertia should be located in priority.
The fact that VSGs do not improve the RoCoF performance for central faults suggests that performances are better when VSGs
are distributed in peripheral regions which support inter-area oscillations~\cite{Rogers2012,fritzschLongWavelengthCoherency2022}, and 
where disturbances are larger to start with~\cite{tylooKeyPlayerProblem2019}.
To test this hypothesis, we compare performances for three different distributions of 30 VSGs. The first two have VSGs  distributed either centrally or peripherally.
They are obtained by picking every other generator out of the 60 most or least central ones, respectively,
according to their average resistance distance centrality~\cite{Klein1993}. The third one considers 30 VSGs homogeneously distributed in terms of their centrality. 
The location of the VSGs in each case is shown in Fig.~\ref{fig:resistance_map}. To ensure legitimate comparisons, we fixed 
the total minimum amount of inertia in the grid at the same value in all cases.

\begin{figure}
    \centering
    \includegraphics[width=\columnwidth]{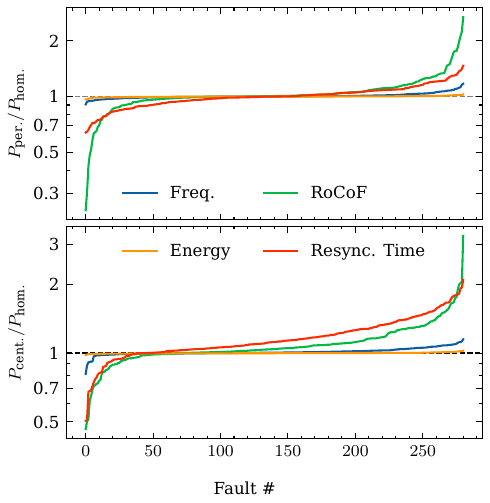}
    \caption{
    Ratio of the performance measures for 30 VSGs distributed in the peripheral (top) and central (bottom) parts of the grid with those obtained 
    for a homogeneous distribution of VSGs (see discussion in the main text and Fig.~\ref{fig:resistance_map}). 
    In the top panel, values below the dashed line correspond to the peripheral distribution performing better and values above correspond to the homogeneous distribution performing better.
    Similarly, in the bottom panel, values below the dashed line correspond to the central distribution performing better and values above correspond to the homogeneous distribution performing better.
    To improve readability, the faults are ordered independently for each curve to make the latter monotonically increasing.\label{fig:remote-homog}}
\end{figure}

The results are presented in Fig.~\ref{fig:remote-homog}. First, we clearly see that the central distribution worsens the grid response according to all performance measures, in almost
all fault cases, especially from the point of view of the RoCoF and resynchronization performance measures. Second, we find that, overall, peripheral and homogeneous geographical distributions of VSGs perform similarly, according to the frequency, the energy and 
the resynchronization time performance measures. Third, the peripheral distribution of VSGs improves the RoCoF performance measure in the majority of fault cases over  
the homogeneous distribution. Furthermore, we find that in almost all instances where the peripheral distribution gives a worse RoCoF response than the 
homogeneous distribution, the RoCoF response itself is small, so that the worsened response is unproblematic. 

Our results therefore indicate that both homogeneous and peripheral distributions of VSGs are viable options, with a slight, but not too significant advantage for the peripheral distribution.
Interestingly, the latter significantly outperforms all other distributions  when the fault is located in the Balkans or the Iberian Peninsula.
These are  the areas supporting the modes driving the east-west inter-area oscillations~\cite{fritzschLongWavelengthCoherency2022}, which
corroborate the hypothesis made in the previous paragraph, that the adaptive inertia control scheme also improves the damping of inter-area oscillations.

\subsubsection{Future scenario with reduced electromechanical inertia}

\begin{figure}
    \centering
    \includegraphics{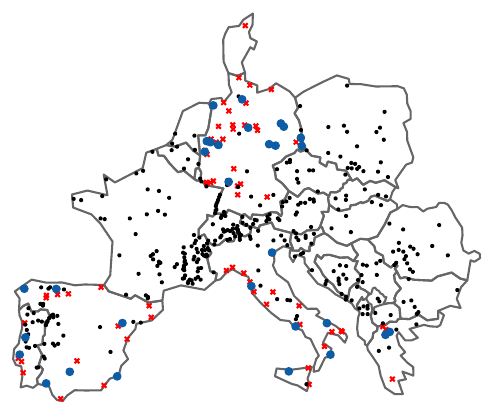}
    \caption{Location of the machines with adaptive inertia (blue dots) and the conventional machines (black dots) for a future scenario in which generators powered by fossil fuels have been replaced with renewable (inertia-less) generation.
    The generators which are turned into renewable generation and not replaced with VSGs are shown by red crosses.
    Note the reduced density of generators in Portugal, Spain, Italy, Greece, Germany, and Denmark.\label{fig:reduced_map}}
\end{figure}

We finally investigate a future scenario of a well-developed energy transition, where the penetration of renewable energy sources has been significantly increased.
Accordingly, we turn all the fossil-fuel generators in Greece, Italy, Germany, Denmark, and the Iberian Peninsula into renewable generators
of the same rated power, i.e. by turning their dynamics from  Eq.~\eqref{eq:swinggen} to Eq.~\eqref{eq:swingload}, while keeping the damping parameter the same.
This corresponds to droop-controlled inverters providing frequency control~\cite{Simpson2012}. In total 94 generators are changed, corresponding to a total inertia 
reduction by \SI{31}{\percent}. They are indicated by red crosses in Fig.~\ref{fig:reduced_map}.

To investigate whether adaptive inertia can compensate for the inertia reduction,
we consider three different distributions of VSGs. First and second, we consider VSGs on  
the peripheral and homogeneous distributions, whose excellent performances have been emphasized in the previous section.
We further take account of the fact that rotating machines can be converted into synthetic inertia as power electronic controlled synchronous condensers.
Accordingly, we consider a third distribution where 30 of the fossil-fuel generators turned renewable generators are promoted to VSGs with adaptive inertia.
This new distribution of inertial generation is shown in Fig.~\ref{fig:reduced_map}. 
To ensure that the minimum amount of inertia is the same in all three cases, we set 
the minimum inertia to 10~\% of the inertia of the former conventional generation for the VSGs installed at renewable buses, and to 
 110~\% of the constant inertia for VSGs installed at buses with conventional generation.

\begin{figure}
    \centering
    \includegraphics{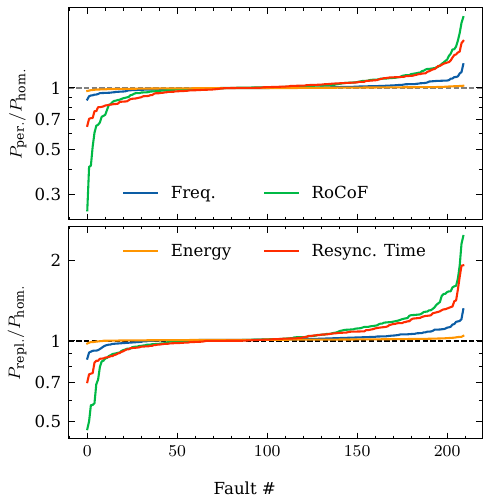}
    \caption{Ratio of the performance measures for 30 VSGs distributed in the peripheral (top) part of the grid and over removed conventional generators (bottom), with those obtained for a homogeneous distribution for the new renewables scenario. 
    Values below the dashed line correspond to our adaptive scheme performing better and values above correspond to the default case with only constant inertia performing better.
    To improve readability, the faults are ordered independently for each curve, to make the latter monotonically increasing.\label{fig:comp_renew}}
\end{figure}

\begin{figure}
    \centering
    \includegraphics{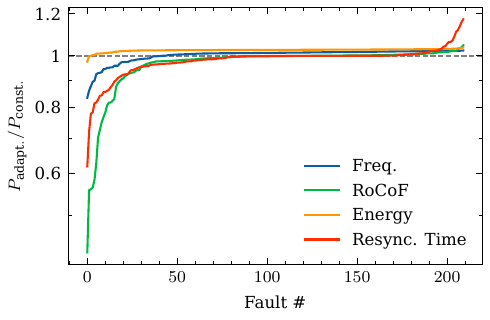}
    \caption{Ratio of the performance measures for 30 VSGs distributed over the removed conventional generators to the ones of the case without VSGs for the new renewables scenario. Values below the dashed line correspond to our adaptive scheme performing better and values above correspond to the default case with only constant inertia performing better. To improve readability, the faults are ordered independently for each curve to make the latter monotonically increasing.\label{fig:comp_reduced}}
\end{figure}

The results are shown in Figs.~\ref{fig:comp_renew} and~\ref{fig:comp_reduced}.
Fig.~\ref{fig:comp_renew} confirms the results of the previous section: the peripheral distribution of VSGs has a slight, though still significant advantage over the homogenous one. They both perform better than the case where VSGs are distributed over fossil-fuel generators turned renewables.
We stress that this latter distribution still contains buses over a large range of centralities.
Still this latter distribution performs significantly better than turning fossil-fuel generators into renewables without inertia compensation, as is 
clearly shown in Fig.~\ref{fig:comp_reduced}. There are very few faults for which the resynchronization time increases a bit, which however correspond to 
the shortest resynchronization times.
We conclude that VSGs with our adaptive inertia schemes are solutions of choice to support grid stability upon increasing the penetration of renewable generation.

\section{Conclusion\label{sec:conclusions}}

Decarbonization of the energy sector requires a strong increase in the penetration of new renewable sources of energy. This increase will be 
accompanied by a significant reduction of available electromechanical inertia. Therefore, 
guaranteeing the dynamical stability of future electric power grids is one of the main challenges facing the energy transition. 
Motivated by the ambivalent nature of inertia, which absorbs frequency disturbances at short times but helps them propagate at later times, 
we propose to deploy virtual synchronous generators controlled by 
an innovative adaptive inertia scheme, Eq.~\eqref{eq:adaptinertia}, to guarantee the stability of future power grids. 
Unlike previously proposed virtual inertia schemes, ours is always stable. Moreover, it outperforms the electromechanical
inertia from conventional generators for both short- and long-time effects, essentially regardless of the location of the considered fault.
As a matter of fact, our scheme not only absorbs strong frequency disturbances at short time scales, but it moreover efficiently damps 
long-range, long-time inter-area oscillations. 
While evidences suggest that VSGs with adaptive inertia should be located in peripheral areas in priority, distributing them homogeneously 
or where fossil-fueled generators have been replaced by NREs also significantly enhances grid stability.
VSGs controlled with the proposed adaptive inertia scheme should therefore be considered as serious candidates to help 
stabilize future decarbonized power grids.

\section*{acknowledgments}
We thank the Swiss National Science Foundation for financial support under grant 200020\_182050.
We thank Mar\'ia Mart\'inez-Barbeito, Pere Colet, Florian D\"orfler, Dami\`a Gomila, and Laurent Pagnier for discussions.

%

\clearpage
\onecolumngrid
\begin{center}
\textbf{\large Stabilizing Large-Scale Electric Power Grids with Adaptive Inertia: Supplemental Material}
\end{center}
\setcounter{equation}{0}
\setcounter{figure}{0}
\setcounter{table}{0}
\setcounter{section}{0}
\makeatletter
\renewcommand{\theequation}{S\arabic{equation}}
\renewcommand{\thefigure}{S\arabic{figure}}
\renewcommand{\bibnumfmt}[1]{[S#1]}
\renewcommand{\citenumfont}[1]{S#1}

\title{Stabilizing Large-Scale Electric Power Grids with Adaptive Inertia: Supplemental Material}

\author{Julian Fritzsch}
\author{Philippe Jacquod}
\affiliation{Department of Quantum Matter Physics, University of Geneva, CH-1211 Geneva, Switzerland}
\affiliation{School of Engineering, University of Applied Sciences of Western Switzerland HES-SO, CH-1950 Sion}
\date{\today}
\maketitle

\section{Single Machine Infinite Bus Model}
\begin{figure}[H]
    \centering
    \includegraphics[width=.6\textwidth]{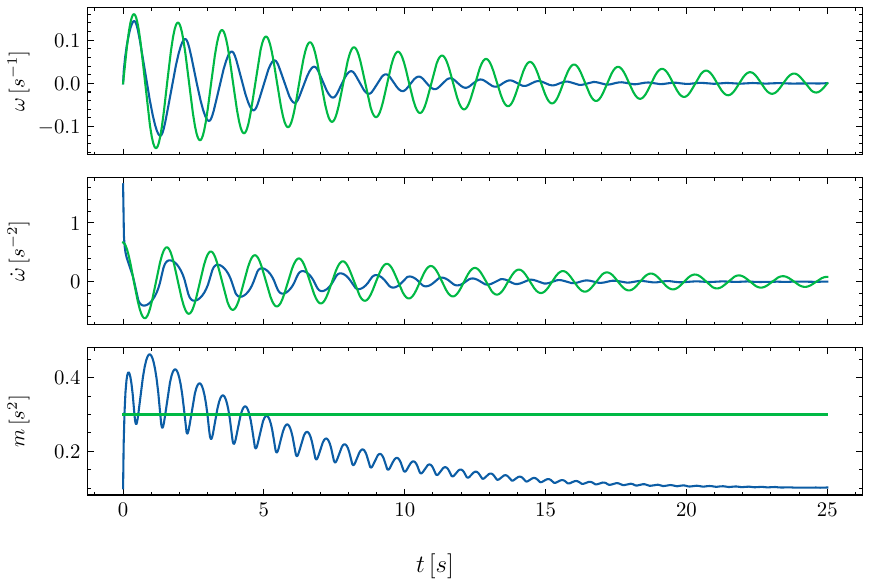}
    \caption{Frequency (top), RoCoF (middle), and inertia (bottom) of the SMIB model following a change in power.
    The blue lines correspond to the model equipped with our adaptive inertia, the green lines to the case with constant inertia.\label{fig:smib}}
\end{figure}

In this section, we demonstrate our adaptive inertia method using a single machine infinite bus (SMIB) model.
This corresponds to a generator being connected to a bus whose voltage angle is fixed at $\theta = 0$ and that can absorb an infinite amount of power.
The swing equations~(1) reduce to 
\begin{equation}\label{eq:smib}
    m \dot{\omega} + d \omega = P - b\sin(\theta),
\end{equation}
where $m$, $d$, and $P$ are the inertia, damping, and power of the generator.
$\theta$ and $\omega = \dot{\theta}$ are its voltage angle and frequency deviation.
$b$ is the susceptance of the line connecting the generator to the infinite bus.
Equipped with our adaptive inertia method, Eq.~\eqref{eq:smib} is extended by
\begin{equation}
    \dot{m} = \alpha\lvert\dot{\omega}\rvert - \beta(m - m_\mathrm{min}).
\end{equation}
Fig.~\ref{fig:smib} shows the reaction of the generator to a change in the power output.
Following the large initial RoCoF (middle panel) the inertia quickly rises (bottom panel) to quickly mitigate the RoCoF.
After the RoCoF has been sufficiently reduced, the inertia decays and returns to its minimal value.
The frequency of the generator with the adaptive inertia is already synchronized when the frequency of the generator with constant inertia is still oscillating (top panel).

\section{Additional Parameter Sweeps for IEEE RTS-96 Network}
\begin{figure}
    \centering
    \includegraphics[width=.6\textwidth]{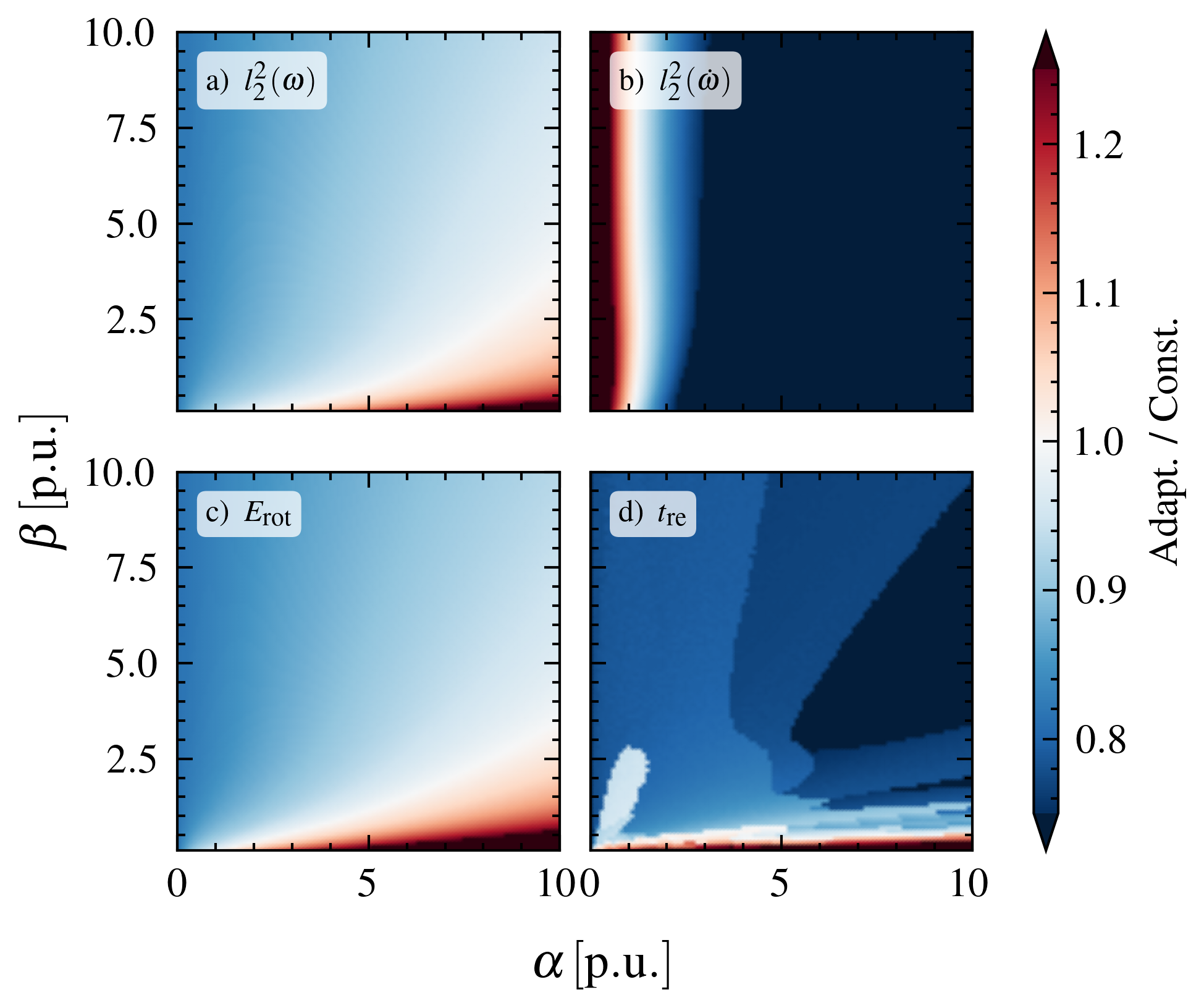}
    \caption{Dependence of a) the frequency performance measure, Eq.~(8), b) the RoCoF performance measure, Eq.~(9), c) the injected inertial energy, Eq.~(10), and d) the resynchronization time, tre, on the VSG control parameters $\alpha$ and $\beta$ defined in Eq.~(3). The fault considered is a change in the power at the VSG on the left below the one indicated by the black arrow in Fig. 2. Color coded are the ratio of the performance measures with VSGs over those with only conventional generation. Blue colored areas correspond to the adaptive method performing better than conventional generators.\label{fig:sweepvsg}}
\end{figure}

To show the generality of the results of Sec. IV A of the original paper, we repeat the numerics of Fig. 3 for the second VSG in Area 1 (located to the bottom left in Fig. 2).
The results are shown in Fig.~\ref{fig:sweepvsg}.
It is clear that all the observations made in Sec. IV still hold.

\begin{figure}
    \centering
    \includegraphics[width=.6\textwidth]{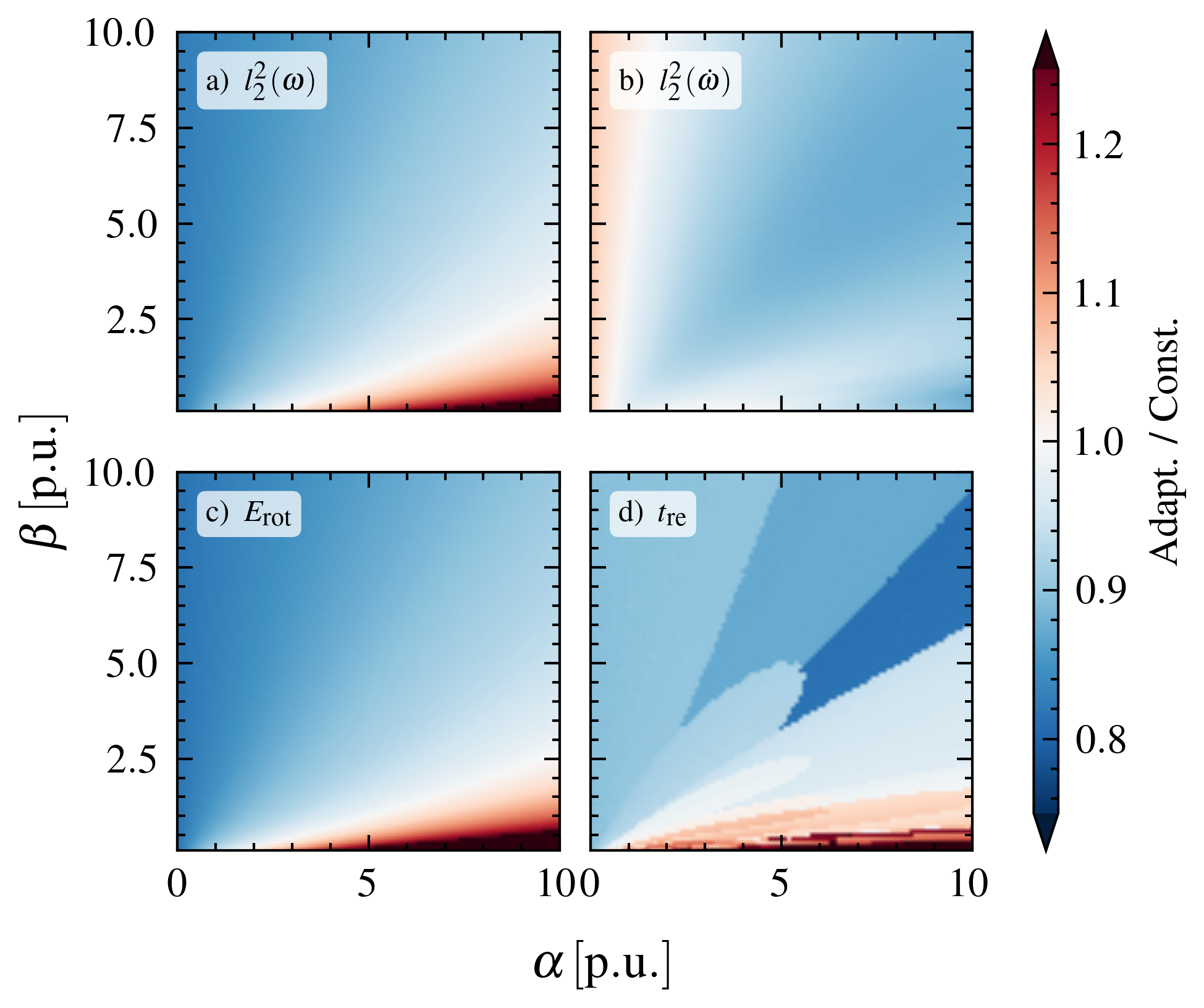}
    \caption{Dependence of a) the frequency performance measure, Eq.~(8), b) the RoCoF performance measure, Eq.~(9), c) the injected inertial energy, Eq.~(10), and d) the resynchronization time, tre, on the VSG control parameters $\alpha$ and $\beta$ defined in Eq.~(3). The fault considered is a change in the power at the generator indicated by the red arrow in Fig. 2. Color coded are the ratio of the performance measures with VSGs over those with only conventional generation. Blue colored areas correspond to the adaptive method performing better than conventional generators.\label{fig:sweepconst}}
\end{figure}

Finally, we also show the performance measures for a fault on the constant generator indicated by the red arrow in Fig. 2 in Fig.~\ref{fig:sweepconst}.
We find that the general observations still hold, even though the impact on the RoCoF performance is smaller.
This is not surprising as the largest contributor to the RoCoF performance measure is the faulted generator that in this case is not equipped with our adaptive inertia method.

\end{document}